\documentclass[useAMS,usenatbib]{mn2e}

\usepackage{natbib}
\usepackage{graphicx}

\title[Polarization of Directly Imaged Exoplanets]{Probing the  
Physical Properties of Directly Imaged Gas Giant Exoplanets Through  
Polarization}

\author[M. S. Marley and S. Sengupta]{Mark S. Marley$^{1}$\thanks{E- 
mail: Mark.S.Marley@NASA.gov} and Sujan Sengupta$^{2}$\thanks{E-mail:
sujan@iiap.res.in}\\
$^{1}$NASA Ames Research Center, MS-245-3, Moffett Field, CA 94035,
U.S.A.\\
$^{2}$Indian Institute of Astrophysics, Koramangala 2nd Block,
Bangalore 560 034, India\\}

\begin{document}

\date{Submitted 2011 March 10.}

\pagerange{\pageref{xx}--\pageref{yy}} \pubyear{2011}

\maketitle

\label{firstpage}

\begin{abstract}
It is becoming clear that the atmospheres of the young, self-luminous extrasolar giant planets  
imaged to date are dusty.
Planets with dusty atmospheres may exhibit detectable amounts of linear polarization in  
the near-infrared, as has been observed from some field L dwarfs. The asymmetry required in the thermal radiation field to produce polarization 
may arise either from the rotation-induced oblateness or from surface
inhomogeneities, such as partial cloudiness.  While it is not possible at present to  predict
the extent to which atmospheric dynamics on a given planet may produce surface inhomogeneities  substantial enough to
produce net non-zero disk integrated polarization, the contribution of
rotation-induced oblateness can be estimated. Using a self-consistent, spatially homogeneous  
atmospheric model
and a multiple scattering polarization formalism for this class of
exoplanets, we show that polarization on the order of 1\% may arise due to
the rotation-induced oblateness of the planets.  The degree of polarization for cloudy planets should peak
at the same wavelengths at which the planets are brightest in the near-infrared.  The observed polarization may be even 
higher if surface inhomogeneities  
exist and play a significant role. Polarized radiation from self-luminous gas giant exoplanets, if detected,  
provides an additional tool to characterize these young planets and a  
new method to
constrain their surface gravity and masses.

\end{abstract}

\section{Introduction}

Several young, self-luminous gas giant planets have been
detected by direct imaging \citep{Cha04, Mar08, Mar10, Lag10, Laf10} around  
nearby stars. These
objects are now being characterized by photometry and even  
spectroscopy \citep{Bow10, Pat10, Cur11,Bar11} in
an attempt to characterize their atmospheres and constrain the  
planetary masses. In the next
few years many more such planets are almost certainly to be detected by
ground-based adaptive optics coronagraphs, such as the P1640 coronagraph
on Palomar, the Gemini Planet Imager, and SPHERE on the VLT  
\citep{Bei10}.

The characterization of the mass of a given directly imaged planet can  
be
problematical, since such planets typically lie at large star-planet
separations (tens of AU and greater) and are thus not amenable to  
detection
by radial-velocity methods. Instead masses must be estimated either by
comparison of photometry and spectroscopy to planetary evolutionary and
atmospheric models or by their gravitational influence on other planets or disk \citep[e.g.][]{Kal05,Fab10}.
Model comparisons as a method for constraining mass can be ambiguous,
however.  Evolution models which predict luminosity as a function of age
have yet to be fully tested in this mass range for young planets and at very young ages ($< 100$ Myr)
the model luminosity can depend on the unknown initial conditions \citep{Mar07, For08}.  
The masses of the planets around HR 8799
estimated by cooling models are apparently inconsistent with standard model spectra \citep{Bow10, Bar11, Cur11} and can
 lead to rapid orbital instabilities if
circular, face-on orbits are assumed \citep{Fab10}.  Finally the  
mass of the
planetary mass companion to the brown dwarf 2M1207 b \citep{Cha04}
inferred from fitting of spectral  models to observed near-infrared  
colors is discrepant with the mass inferred from the companion's  
luminosity and the age of the primary \citep{Moh07}. 

Discrepancies such as these may arise because young exoplanets exist  
in a gravity-effective temperature
($g,T_{\rm eff}$) regime in which both the evolutionary and atmospheric models have yet to be  
validated. Fits of photometry and  
spectroscopy
to predictions of atmosphere models depend upon the veracity of the  
models
themselves, which---in the $T_{\rm eff}$ range of interest---in turn
sensitively
depend upon model cloud profiles, which are as yet uncertain.    
Extensive
experience with fitting models to brown dwarf spectra and photometry
\citep{Cus08, Ste09}  reveals that while effective temperature can be  
fairly
tightly constrained, gravity determinations are usually less precise,  uncertain
in some cases by almost an order of magnitude in $g$.  While there are low gravity spectral
indicators recognized from surveys of young objects \citep{cruz09,  
kirk06}
  these have yet to be
calibrated by studies of binary objects which allow independent  
measures of
mass.  Ideally for a single object  with a given radius $R$, evolution  
model
luminosity, which (for a known parallax) constrains $R^2T_{\rm eff}^4$  
would
be fully consistent with $(g, T_{\rm eff})$ constraints from  
atmosphere model
fitting.  But, as noted above, this is often not in fact the case, as the derived luminosity, mass, and radii of the companion to
2M1207 b as well as the HR 8799 planets are not fully internally self-consistent with standard evolution models.

Given the likely future ubiquity of direct detections of young, hot  
Jupiters and the clear need for
additional independent methods to constrain planet properties, we have  
explored the utility
of polarization as an additional method for characterizing self-luminous planets.

Polarization of close-in giant exoplanets whose hot atmosphere favours
the presence of silicate condensates, is discussed by \cite{Sea00} and by 
\cite{sengupta}.  While these authors
considered the polarization of the combined light from an unresolved
system of star and planet, \cite{Sta04} presented the
polarization of the reflected light of a resolved, directly-imaged Jupiter-like exoplanet.
Since the polarized light of a close-in
exoplanet is combined with the unpolarized continuum flux of the star
which cannot be resolved, the amount of observable polarization in such
case is extrmeley low -- of the order of magnitude of planet-to-star flux
ratio.  Polarization measurements of directly-imaged exoplanets in reflected light is also challenging. The
removal of scattered light from the primary star must be precise in both polarization
channels so that the planet's intrinsic polarization (which is a differential measurement) can be accurately determined.
In any case no extrasolar planet has yet been imaged in scattered light, an accomplishment that will likely require a 
space-based coronagraph \citep[e.g.,][]{Boc11}.
Measuring polarization of thermally emitted radiation---as we propose here---is also difficult but does not require a planet
to be close to the star (where the starlight suppression is most difficult) so that it is bright in reflected light.  Furthermore
extrasolar planets have already been imaged which raises the possibility of polarization observations.

It is clear from  comparisons of model spectra to data that most of the
exoplanets directly imaged to date have
dusty atmospheres \citep{Mar08,Bow10,Laf10,Bar11,Cur11,Ske11}. Clear atmospheres
lacking dust grains can be polarized, but only at blue optical  
wavelengths
where gaseous Rayleigh scattering is important \citep{Sen09}.  Since even the  
hottest
young exoplanets will not emit significantly in the blue, grain scattering must be  
present for
there to be measurable polarization in the near-infrared where warm  
giant
planets are bright \citep{Sen10}.  There are two temperature ranges  
within
which we expect a gas giant exoplanet to possess significant atmospheric
condensates.  The first is L-dwarf like planets (roughly $1000 <  
T_{\rm eff}   <  2400\,\rm K$) where iron and silicate grains condense in the  
observable
atmosphere. The lower end of this range in the planetary mass regime is as yet uncertain. 
The second temperature range occurs in cool planets with atmospheric water
clouds ($T_{\rm eff} < 400\,\rm K$).  There have yet been no confirmed  
detections of such planets.  Here we will focus on  
the first category
since such objects are brighter, more easily detectable, and the  
comparison
to the field dwarfs is possible.

Although survey sizes are fairly small, linear polarization of field L  
dwarfs
has been detected.  \cite{Men02} and \cite{Oso05} both report that a  
fraction
of L dwarfs, particularly the later, dustier spectral types, are
intrinsically polarized.  \cite{Sen10} find that the observed  
polarization can plausibly arise from emission of cloudy, oblate dwarfs, although  
to produce the required oblateness (20\% or more) the dwarfs must have  
fairly low gravity for a field dwarf ($g\sim 300\,\rm m\,s^{-2}$) and rapid  
rotation.  The required rotation periods are brisk, as little as 2 hours or less,  
but are compatible with observed rotational velocities in at least some  
cases (see \cite{Sen10} for a discussion).  \cite{Sen10} further find
that the near-infrared polarization is greatest at $T_{\rm eff} \sim 1600\,\rm K$ where
their model condensate clouds are both optically thick and still prominent enough
in the photosphere to maximally affect the polarization.

Surface inhomogeneities can also give rise to a net polarization \citep{Men04} and experience
from the solar system confirms that irregularly spaced clouds are to be expected.  
Both Jupiter's and Saturn's thermal emission
in the five-micron spectral window is strongly modulated by horizontally
inhomogeneous cloud cover and it would not be surprising to find
similar morphology in the atmospheres of exoplanets.
In the presence
of surface inhomogeneity, the asymmetry that produces the net non-zero
disk-integrated polarization would increase and hence a combination of  
oblate
photosphere and surface inhomogeneity can give rise to detectable  
levels of polarization.

Exoplanets are even better candidates than L dwarfs to have an oblate  
shape
and be polarized.  With a lower mass and roughly the same radius as a  
brown
dwarf (and thus a lower gravity), a rapidly rotating planet can be
significantly oblate and consequently produce  a polarization
signal even without surface inhomogeneities.
Here we explore the conditions under which  the thermal emission from  
a warm,
young exoplanet may be polarized and consider the scientific value
of measuring exoplanet polarization.  We first look at the issue of  
oblateness,
then present a set of cloudy model atmosphere calculations relevant to planets
amenable to direct detection and discuss under which conditions their  
thermal
emission may be polarized.  Finally we discuss our findings and  
explore how
the characterization of an extrasolar planet may be enhanced by
polarization observations.

\section{Young Giant Exoplanets}
\subsection{Evolution}

When giant planets are young they are thermally expanded and boast
larger radii and smaller gravity.  \cite{For08} have computed evolution
models for gas giant planets with masses ranging from 1 to $10\,\rm M_J$
for ages exceeding $10^6\,\rm yr$ and we can use their results to  
predict
the oblateness of thermally expanded young Jupiters with various  
rotation rates. Those authors modeled two types of evolution models.  
The first variety, termed `hot starts', was most  
traditional
and assumed the planets formed from hot, distended envelopes of gas  
which
rapidly collapsed.  This calculation is comparable to that of most other
workers in the field.  They also presented calculations for planets fomed
by  the core accretion planet formation process (see \cite{Lis07}) which  
(depending
on details of the assumed boundary condition for the accretion shock)
produces planets that are initially much smaller and cooler than in the
`hot start' scenario.

For the calculations here we choose to  use the `hot start'
evolutionary calculation.  We do this for several reasons.  First, these
models provide a reasonable upper limit to the radius at young ages and
thus bound the problem.  Second, at the large orbital separations that
will, at least initially, be probed by ground based adaptive optics
coronagraphic imaging, the core accretion mechanism may be inefficient
at forming planets.  Thus the gaseous collapse scenario may be more
relevant choice.  Finally the three planets observed around HR 8799
are all much brighter than predicted by the \cite{For08} cold-start, but
not the hot-start, cooling tracks.

Figure 1 presents model evolution tracks for non-irradiated giant  
exoplanets from \cite{For08}.  On this figure planets age from the  
right to the left as effective temperature falls, the planets  
contract, and their surface gravity, $g$, increases.  The dashed lines  
denote isochrones.  This figure guides our selection of atmosphere  
models to evaluate for polarization studies.  Groundbased
coronagraphic searches for planets are expected to foucus on stars
younger than about 200 Myr \citep[e.g.,][]{McB11}.  From the figure we see that  at ages of 10 to 200 Myr  
we expect exoplanets with masses falling between 1 and $10\,\rm M_J$  
to have $g$ roughly in the range of 15 to $200\,\rm m\,s^{-2}$.

\subsection{Shape}

Both Jupiter and Saturn are oblate.  The fractional difference,
$f=1-R_p/R_e$,  
between their equatorial and polar radii, known as oblateness, are
0.065 and 0.11 respectively.  The  
extent to which their equators bulge outwards depends on their surface gravity, $g$, and  
rotation rate, $\Omega$, as well as their internal distribution of mass.    
The Darwin-Radau relationship \citep{Bar03}
connects these quantities of interest:
\begin{eqnarray}\label{obl1}
f=\frac{\Omega^2R_e}{g}\left[\frac{5}{2}\left(1-\frac{3K}{2}\right)^2+ 
\frac{2}{5}\right]^{-1}
\end{eqnarray}
Here $K=I/(MR_e^2)$, $I$ is the moment of inertia of the spherical 
configuration, and $M$ and $R_e$ are the mass and equatorial radii. 

The relationship for the oblateness $f$ of a stable
polytropic gas configuration under hydrostatic equilibrium is also derived
by \cite{cha33} and can be written as 
\begin{eqnarray}\label{obl2}
f=\frac{2}{3}C\frac{\Omega^2R_e}{g}
\end{eqnarray}
where $C$ is a constant whose value depends on the polytropic index.

The above two relationships provide the same
value of oblateness for any polytropic configuration. Equating  Eq.~(1) 
and Eq.~(2), we obtain 
\begin{eqnarray}
C=\frac{3}{2}\left[\frac{5}{2}\left(1-\frac{3K}{2}\right)^2+ \frac{2}{5}\right].
\end{eqnarray}
Substituting the value of $K$ for a polytrope of index $n$ gives the value
of the corresponding $C$. For example, $K=0.4, 0.261, 0.205, 0.155, 0.0754 $ for $n=0,1,1.5,2,3$ respectively.
The corresponding values of $C$ derived by \cite{cha33}(p. 553, Table 1).
are 1.875, 1,1399, 0.9669, 0.8612, 0.7716 respectively.

The interiors of gas giant planets can be well approximated as $n=1$ polytropes.
For the observed mass, 
equitorial radii, and rotation rates of Jupiter and Saturn,
expression (2) predicts, with $n=1$, an oblateness 
of 0.064 and 0.11, in excellent  agreement with the observed values.
Figure 2 presents the oblateness computed employing Eq.~(2) as applied  
to 1
and $10\,\rm M_J$ planets at three different ages, 10, 100, and 1,000  
Myr
using the \cite{For08} hot-start cooling tracks.  Also shown is the
oblateness (0.44) at which a uniformly rotating  $n =1.0 $ polytrope
becomes unstable \citep{Jam64}.  Clearly for rotation rates comparable
to those seen among solar system planets we can expect a substantial  
degree
($f>0.10$) of rotational flattening.  As gas giants age and contract the
same rotation rate produces much less oblate planets. However for young,
Jupiter mass planets rotation rates of 7 to 10 hours can
easily produce $f\sim 0.2$ even for planets as old as 100 Myr.  More
rapid rotation rates may produce even greater degrees of flattening.   
L dwarfs, with much higher surface gravity, must have even more rapid  
rotation rates to exhibit even modest flattening \citep{Sen10}.

\section{Polarization of Young Exoplanets}

To explore the degree of polarization expected for various planet masses and ages 
we
considered a selection of one-dimensional, plane-parallel, hydrostatic,
non-gray, radiative-convective equilibrium atmosphere models with sixty vertical layers 
\citep{Ack01, Mar02, Freed08} for specified effective temperatures, 
$800< T_{\rm  eff} < 1200 \, \rm K$ and surface gravities $g = 30$ and 
$100\,\rm m \,sec^{-2}$.
We focus on this apparently limited parameter range since all gas  
giant exoplanets with masses below $10\,\rm M_J$
will have cooled below 1200 K by an age of 30 Myr (see Figure 1 and also \cite{For08}).
The median age for nearby ($< 75\,\rm pc$) young stars that are likely targets for planet imaging surveys
is $50\,\rm Myr$ \citep{McB11}.   For our study we choose a lower limit of  
800 K, well below the $T_{\rm eff}$ at which Sengupta \& Marley (2010) predicted
maximal polarization for field L dwarfs.  At such temperatures dust clouds, if present globally across the disk, will lie at high
optical depth and we expect produce a smaller polarization signal than the warmer objects.  

Indeed 800 K is well below the field dwarf L to T transition temperature of 1200 to 1400 K (\cite{Ste09} and references therein) by which point most signs of clouds have departed.  However
there exists growing evidence that there is a gravity dependence to the
effective temperature at which clouds are lost from the atmosphere and
certainly the planets such as those orbiting HR 8799 are still dusty at
effective temperatures near 1000 K \citep{Bow10}. 
  Observation of a polarization signal in a cooler exoplanet would provide  
powerful evidence for atmospheric dust.

Some of the more massive young exoplanets
($M>8\,\rm M_J$) may have gravities in excess of our $100\,\rm m 
\,sec^{-2}$ upper limit, but as we show below
little oblateness-induced polarization is expected at high gravity in  
this $T_{\rm eff}$ range (see also \citet{Sen10}).
For example a surface gravity of $g=100\,\rm m\,sec^{-2}$ and $T_{\rm  
eff}= 1000\,\rm K$
approximately describes an $8\,\rm M_J$ planet at an age of 100  
Myr while values of $30\,\rm m\,sec^{-2}$ and 800 K are expected for a
$2\,\rm M_J$ planet at an age of 60 Myr.  We choose these values and a few others to illustrate  
the parameter space and the sensitivity of the results to variations in  
gravity and effective temperature.  

Each model includes atmospheric silicate and
iron clouds computed with sedimentation efficiency \citep{Ack01}  
$f_{\rm sed}=2$.   Preliminary studies by our group suggest that even dustier
models with $f_{\rm sed}\sim 1$ might be necessary to reproduce the HR 8799 planets.  However our previous
work \citep{Sen10} has demonstrated that while $f_{\rm sed}= 1$ atmospheres do show greater
polarization than $f_{\rm sed} = 2$, the difference is slight when integrated over the disk.  Other cloud
modeling approaches are reviewed by \cite{Hel08}.  Some of these alternative cloud modeling formulations,
such as those employed by Helling and collaborators \citep[e.g.,][]{Hel06, Hel08b}, predict a greater abundance of small particles
high in the atmosphere than the Ackerman \& Marley approach.  Such a haze of small particles could potentially produce a larger polarization signal than we derive here.  Polarization measurements
may thus help provide insight into the veracity of various approaches.

 As in \cite{Sen10} we employ the gas and dust opacity, the temperature-pressure profile and the dust scattering asymmetry function averaged over each
atmospheric pressure level derived by the atmosphere code in a multiple
scattering polarization code that solves the radiative transfer  
equations in vector form to calculate the two Stokes parameter $I$ and $Q$ in a  
locally plane-parallel medium \citep{Sen09}.  For each model layer we fit a Henyey-Greenstein phase function to
the particle scattering phase curve predicted by a Mie scattering calculation.  A combined Henyey-Greenstein-Rayleigh phase matrix \citep{Liu06} is then used to calculate the angular  
distribution of the photons before and after scattering.  In the near-infrared the contribution of Rayleigh scattering by the gas to 
the overall scattering is negligible and the scattering is treated in the Henyey-Greenstein limit with the particle
phase function computed from Mie theory.  Specifically the off diagonal terms of the
scattering phase matrix are described by \cite{Whi79} and are very similar to the
pure Rayleigh case.  For the diagonal elements the Henyey-Greenstein elements are
used.  In the limit of the scattering asymmetry parameter approaching zero the matrix converges to the Rayleigh scattering limit.
Finally, the angle dependent  
$I$ and $Q$ are integrated over the rotation-induced oblate disk of the object
by using a spherical harmonic expansion method and the degree of  
polarization is taken as the ratio of the disk integrated polarized flux ($F_Q$) to  
the disk integrated total flux ($F_I$). The detailed formalisms as well as the  
numerical methods are provided in \cite{Sen09}.

Figures 3 and 4 illustrate typical input properties of the models employed here.  Figure 3 shows a model temperature-pressure profile along with iron and silicate condensate grain sizes as computed by our cloud model.  Figure 4 shows the mean layer single scattering albedo, $\overline{\omega_0}$, and scattering asymmetry parameter, $\overline{\cos \theta}$, as a function of wavelength near the peak opacity of the cloud.  Within strong molecular bands the single scattering albedo approaches zero since gas absorption dominates over the cloud opacity.  Below the cloud base and far above the cloud both the albedo and asymmetry parameters are essentially zero in the near infrared as gaseous Rayleigh scattering makes little contribution to the opacity at those wavelengths.  Note that for the computed particle sizes the cloud is strongly forward scattering at wavelengths where cloud opacity dominates molecular absorption in agreement with a recent study by \cite{Dek11}.

\section{Results and Discussions}

Figure~5 presents the computed thermal emission and polarization  
spectra of an approximately 10 Myr old 2 Jupiter mass planet assuming rotation periods
of 5 and 6 hrs.  The striking dependence of polarization on the rotation rate arises from the sensitivity of oblateness to rotation period
as seen in Figure 2.  Generally speaking the degree of polarization is highest at those wavelengths of low gaseous opacity where the cloud is visible  while at other wavelengths, inside of atomic and molecular absorption bands, flux emerges from higher in the atmosphere and is less influenced by cloud scattering.  While the degree of polarization peaks at the shortest
wavelengths shown (from the influence of gaseous Rayleigh scattering), there is very little flux at optical wavelengths.
However in the near-infrared, where windows in the molecular opacity 
allow flux to emerge from within the clouds, the computed  
degree of polarization approaches 1\%.  In these spectral regions the planets  
will be bright, the contrast with the primary star favorable, and thus the
polarization may be more easily detectable at this level.  Beyond about $2.2\,\rm \mu m$ thermal emission emerges from above the cloud tops and thus there is no signature of the scattering and the net polarization is near zero.  This pattern of polarization is diagnostic of atmospheric clouds
and is easily distinguished from other sources of polarization, for example a circumplanetary disk.

Figures 6 and 7 show  warmer model cases for the same gravity with similar behavior.  These cases would apply to quite young planets at an age of less than ten million years, but illustrate that the degree of polarization does not dramatically increase at higher effective temperatures.  Figure 8 shows the variation with gravity.  With a fixed rotation period of five hours, models with $g$ of 56 and $100\,\rm m\,s^{-2}$ show very little polarization at any wavelength.  These models would correspond to approximately 4 to 6 Jupiter mass planets at ages greater than 10 million years, perhaps typical of the planet types that may be directly imaged.  The sensitivity of polarization to gravity seen in this figure illustrates the promise of polarization, in the right circumstances, to provide a new constraint on exoplanet mass.

Figure~9 generalizes these trends, showing  the predicted polarization  in $I$
and $J$ bands as a function of the rotational period $P_{rot}$.
 For a  fixed
surface gravity and viewing angle, $i$, the degree of polarization does not
vary substantially within the range of $\rm T_{eff}$ between 800 and  
1200 K.
The polarization profiles in both bands increase with decreasing  
rotation period
and the polarization is generally greater in $J$ than in $I$ band.
As is the case for brown dwarfs \citep{Sen10}, for a given rotation  
period
the polarization decreases with lower $i$.

All of the cases shown in Figure 9 have an oblateness less than 0.44, the 
stability limit for an $n=1$ polytrope.  For $g=30\,\rm m\,s^{-2}$ the stability
limit is reached at a rotation period of about 4 hours, slightly less than the
lower limit shown on the figure.  Such short rotation periods may in fact
be a natural consequence of giant planet formation in a circumstellar binary
as the angular momentum of accreting gas naturally produces rapid rotation
rates \citep{War10}.

We conclude that a self-luminous gas giant planet--even with a  
homogeneous cloud distribution--will exhibit
notable polarization (greater than a few tenths of percent) in the near
infrared if the planet is (1) cloudy, (2) significantly oblate, and (3)
viewed at a favorable geometry.   An oblate shape is the easiest to obtain  
at low masses and modest rotation rates or higher masses and more rapid
rotation rates.    Higher effective temperatures, which would produce more dust higher in the atmosphere and more polarization \citep{Sen10},
 are generally excluded by the evolution for ages greater than a few million years.  More massive planets, which take longer to cool,
 have higher gravity and thus a smaller oblateness and less polarization (Figures 2 \& 9) for a given rotation rate.  Given these considerations we believe the cases we have presented here are among the more favorable for homogenous cloud cover.  While we have not considered every combination of parameters, the models presented here along with perturbations of those models we have also studied lead us to conclude that uniformly cloudy planets will not present polarization greater than a few percent and polarization is most likely to be found for young, low mass, rapidly rotating planets.

However inhomogeneous cloud cover, which we have not modeled, may also affect the polarization spectra.
Indeed an inhomogeneous distribution of atmospheric dust (e.g., Jupiter-like  
banding) would not be unexpected.  Such banding
may provide further asymmetry \citep{Men04} and hence increase (or even decrease) the net
non-zero polarization.  A non-uniform  cloud distribution may be the mechanism
that underlies the L to T-type transition among brown dwarfs \citep{Ack01,Bur02,Marl10} and 
variability has  been detected in some transition dwarfs  \citep{Art09,Rad10}.  Cloud
patchiness is also observed in images of thermal emission from   
Jupiter and Saturn taken in the M-band (five-micron) spectral region
\citep[e.g.][]{Wes69, Wes74, Ort96, Bai05}, so patchiness
may indeed be common.  Polarization arising from patchy clouds still requires the presence of  some
clouds of course, thus any polarization detection  provides  
information on
the presence of condensates and--by extension--constrains the  
atmospheric temperature.

\section{CONCLUSIONS}

The next decade is expected to witness the discovery of a great many  
self-luminous extrasolar
giant planets \citep{Bei10}.  The masses, atmospheric composition and  
thermal structure of
these planets will be characterized by photometry and spectroscopy.   
For some systems,
other constraints, such as dynamical interactions with dust disks or  
potential instabilities arising
from mutual gravitational interactions \citep[e.g.,][]{Fab10} may  
also contribute.  Here we
demonstrate that measurable linear polarization in $I$ or $J$ bands reveals the presence
of atmospheric condensates, thereby placing limits on atmospheric  
composition and temperature.  Polarization of thermal emission from a homogeneously cloudy planet
is most favored for young, low mass, and rapidly rotating planets.  A diagnostic characteristic of
cloud-induced polarization is that the polarization peaks in the same spectral bandpasses
as the flux from the planet because photons are emerging from within the cloud itself
as opposed to higher in the atmosphere (Figures 5 through 8).

Assuming that our atmospheric and condensate cloud models are  
reasonably accurate, we conclude that
any measured polarization greater than about 1\% likely can be attributed to  
inhomogeneities in the global cloud deck.  While we have not considered every
possible model case, we find that our most favorable plausible cases do not
produce notably greater polarization.  Other cloud models \citep[e.g.,][]{Hel06, Hel08b} which incorporate
more small particles high in the atmosphere may well produce a different
result, thus polarization may help to distinguish such cases.
For a fixed rottion period, the oblateness and thus polarization
increases with decreasing surface gravity.  In such situations polarization may provide
a new constraint on gravity and mass.  However for gravity in excess
of about 50 $\rm m\,s^{-2}$ and for $T_{\rm eff} < 1200\,\rm K$ (corresponding to planet
masses greater than about $4\,\rm M_J$) we do  not
expect detectable amounts of polarization.  Warmer and higher gravity  
 field L dwarfs, can show measurable polarization since such cloud decks are higher in  
the atmosphere.  For directly imaged exoplanets, however, we do not expect to encounter
such high effective temperatures.  For exoplanets with plausible $T_{\rm eff} < 1200\,\rm K$, 
Figure~9 shows that even if the rotation period is as rapid as 4.5 hrs.  
and the
viewing angle is $90^o$ at which the polarization is maximum, the
percentage degree of polarization in thermal emission is not more that a few times of  
$10^{-2}$.

The aim of our study was to better understand the information conveyed by polarization about the properties of extrasolar giant planets directly imaged in their thermal emission.  We have found that in some cases polarization can provide additional constraints on planet mass, atmospheric structure and cloudiness.  Combined with other constraints, polarization adds to our understanding, although there remain ambiguities.  A study of the polarization signature of partly cloudy planets would  yield further insight into the value of polarization measurements for constraining extrasolar giant planet properties.

\section{Acknowledgements}

We thank the anonymous referee for helpful comments that improved the manuscript. 
MSM recognizes the NASA Planetary Atmospheres Program for support of  
this work.

\clearpage
\begin{figure}
\includegraphics[angle=0.0,scale=0.35]{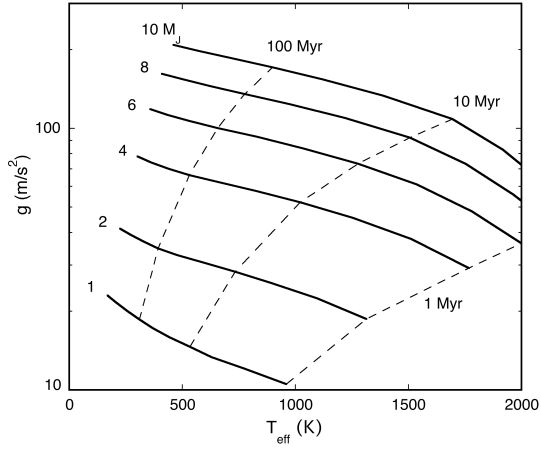}
\caption{Evolution through time in $T_{\rm eff}$ -- $g$ space of non-irradiated, metallicity $[M/H]=0.0$ giant planets of various masses.   
Solid lines are evolution tracks at various fixed masses ranging from  
1 to $10\,\rm M_J$.    Dashed lines are isochrones for various fixed  
ages since an arbitrary `hot start' initial model \citep{For08}.
\label{fig1}}
\end{figure}

\begin{figure}
\includegraphics[angle=0.0,scale=0.35]{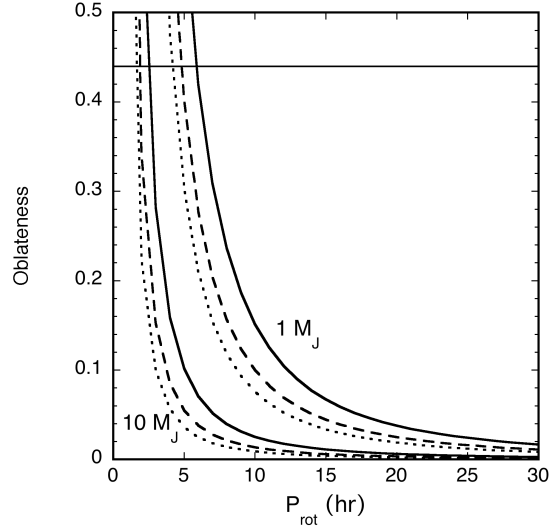}
\caption{Rotationally induced oblateness as a function of rotational  
period for 1 and $10\,\rm M_J$ planets with three different ages  
(10 - solid, 100 - dashed, and 1000 Myr - dotted).  Horizontal line is  
the stability limit for $n=1$ polytropes assuming solid body  
rotation.  More rapidly rotating planets would form a triaxial  
ellipsoidal shape and eventually bifurcate.
\label{fig2}}
\end{figure}

\clearpage
\begin{figure}
\includegraphics[angle=0.0,scale=0.35]{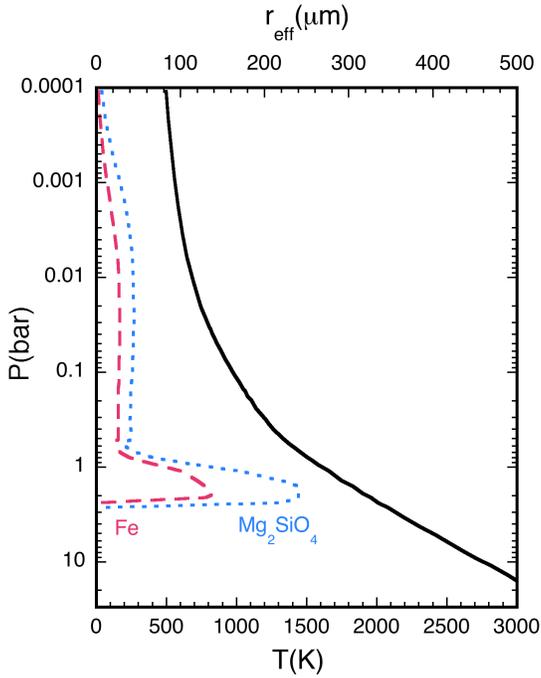}
\caption{Atmospheric temperature ($T$, lower scale) and cloud particle size, $r_{\rm eff}$, (upper scale) as a function of pressure, $P$, for one adopted model atmosphere with $T_{\rm eff} = 1000\,\rm K$, $g=30\,\rm m\,s^{-2}$, and $f_{\rm sed}=2$.  From Figure 2 this corresponds to approximately a $2.5\,\rm M_J$ planet at an age of a few million years. Particle sizes are shown for iron and silicate (forsterite) grains, the two most significant contributors to cloud opacity.  Shown is the ``effective radius'' ($r_{\rm eff}$) of the particles, a mean size precisely defined in Ackerman \& Marley (2001).
\label{fig3}}
\end{figure}

\begin{figure}
\includegraphics[angle=0.0,scale=0.4]{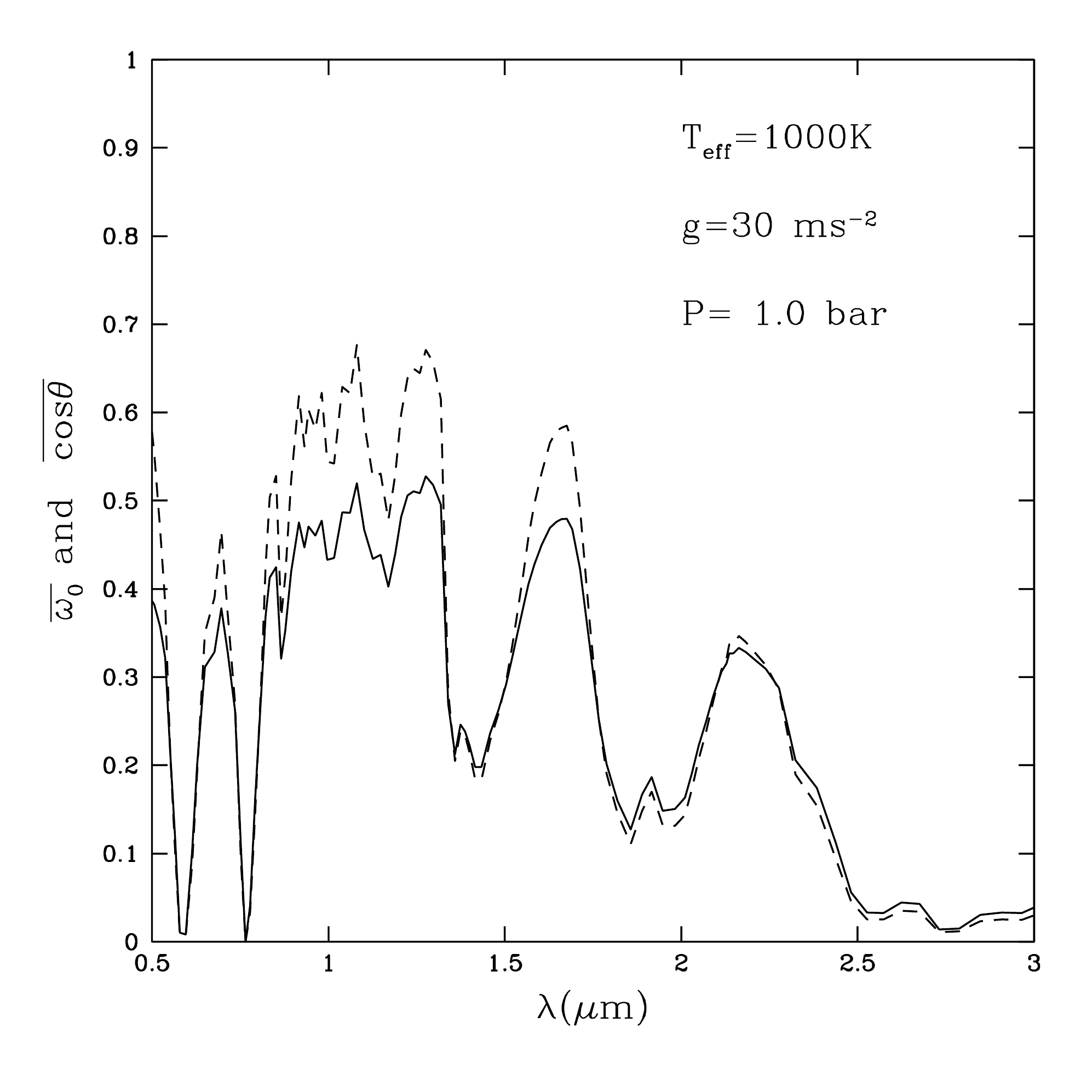}
\caption{Scattering properties as a function of wavelength, $\lambda$, of a model layer near the 1 bar pressure level for the model atmosphere shown in Figure 3.  Shown are the layer single scattering albedo, $\overline{\omega_0}$, solid, and the layer asymmetry parameter, $\overline{\cos \theta}$, dashed.  In strong molecular bands gaseous absorption dominates over scattering, thus lowering the mean layer albedo.
\label{fig4}}
\end{figure}

\begin{figure}
\includegraphics[angle=0.0,scale=.40]{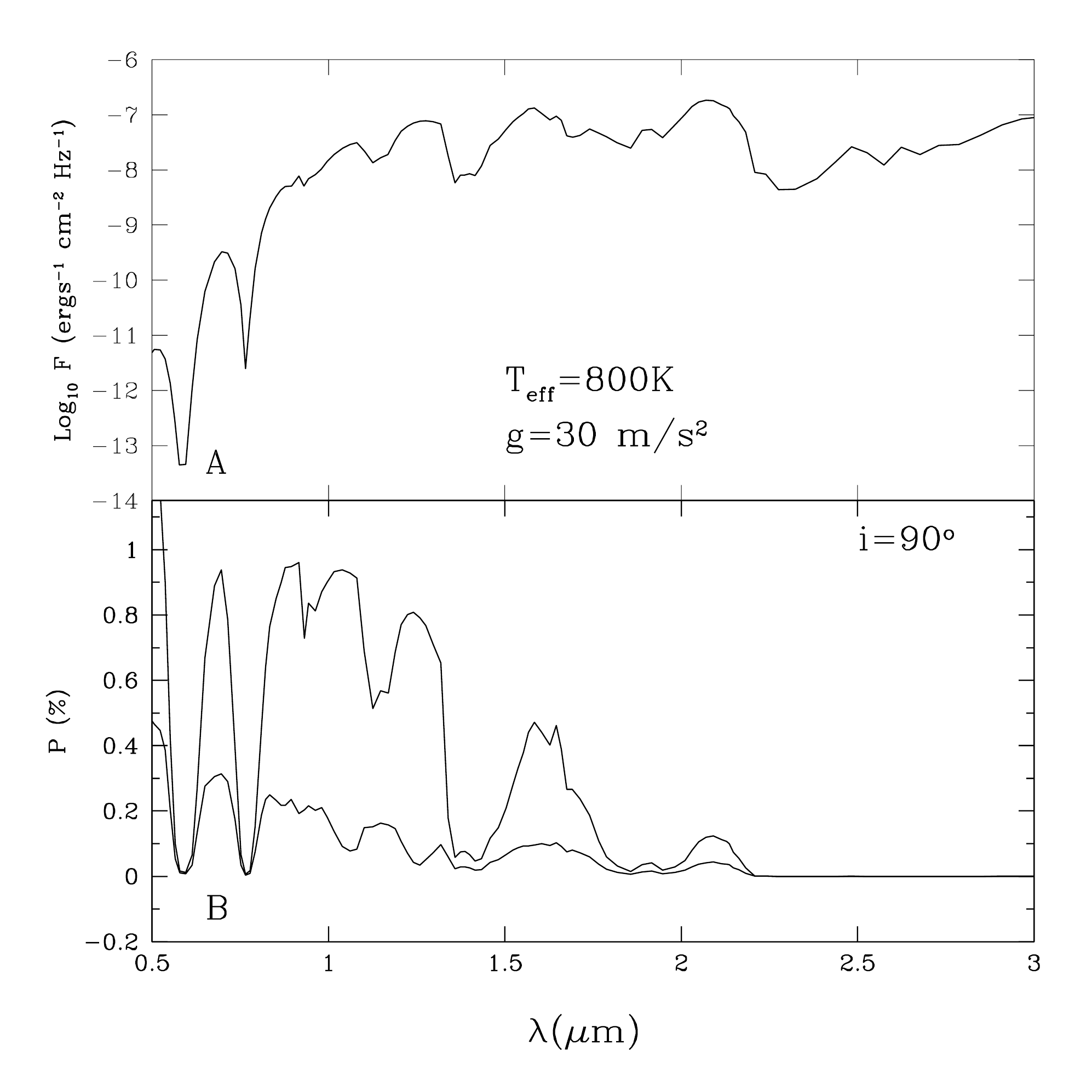}
\caption{The emergent flux (A) and the disk-integrated degree of linear
polarization $P(\%) $ (B) of non-irradiated exoplanets at different wavelengths at viewing angle $i=90^\circ$ (equatorial view). In
(B), the top solid line represents the polarization profile for a
rotational period $P_{\rm rot}= 5\, \rm hr$ while the bottom solid line represents that
for 6 hr.  Note that while the polarization can be high at blue wavelengths there is very little flux there.
\label{fig5}}
\end{figure}

\begin{figure}
\includegraphics[angle=0.0,scale=.40]{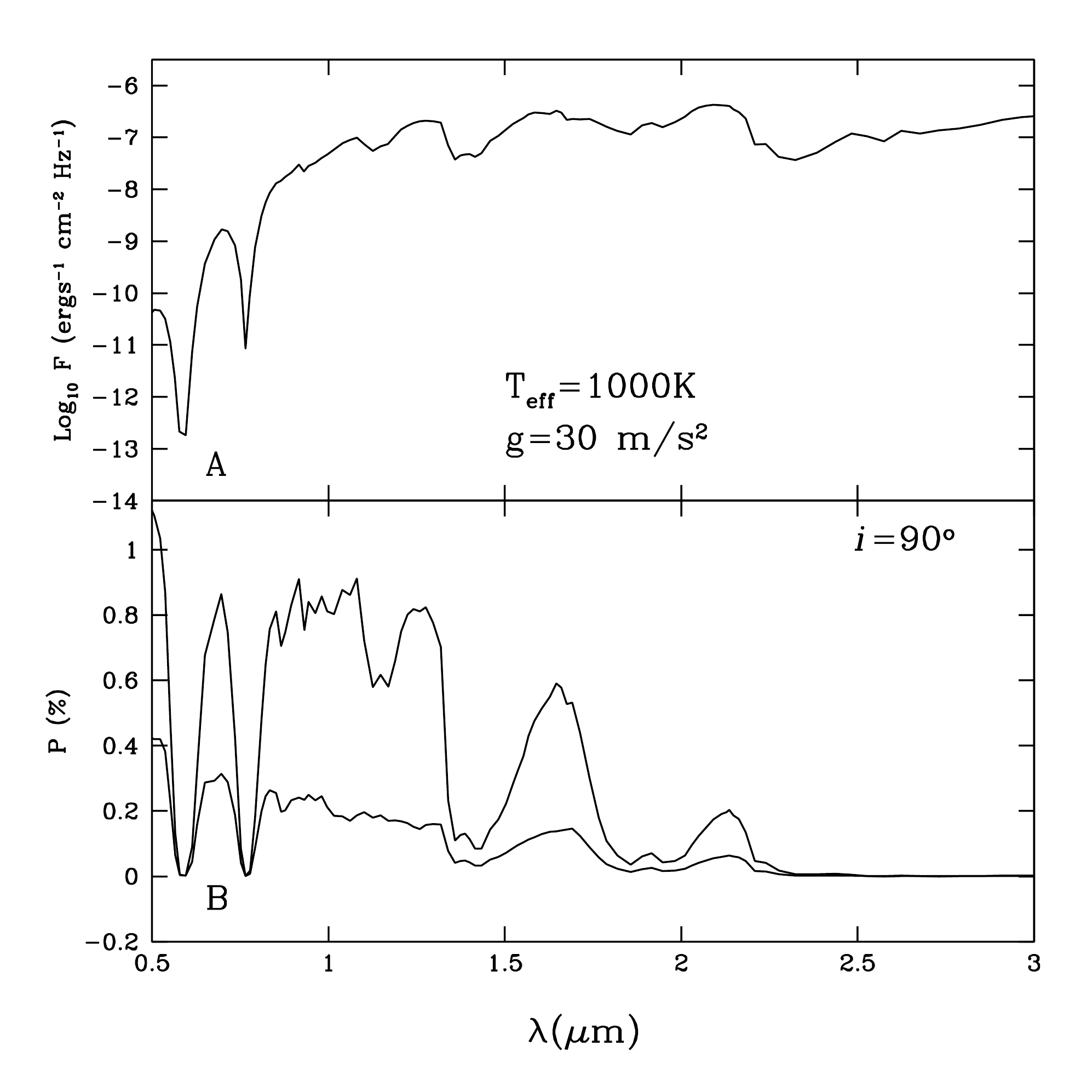}
\caption{Same as figure 5 but with $T_{\rm eff}=1000K$.  This is the result for the model characterized in Figures 3 and 4.
\label{fig6}}
\end{figure}

\begin{figure}
\includegraphics[angle=0.0,scale=.40]{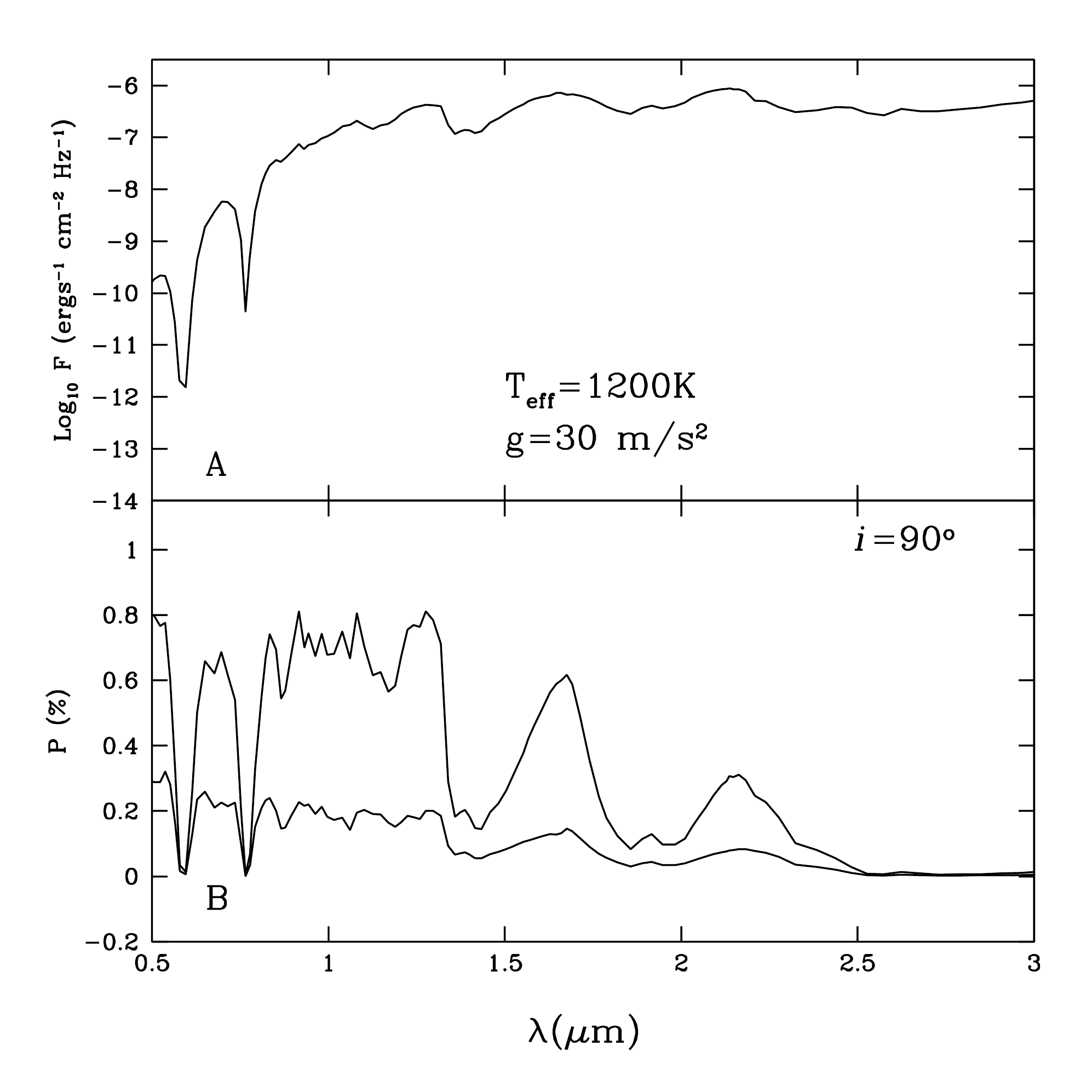}
\caption{Same as figure 5 but with $T_{\rm eff}=1200K$.
\label{fig7}}
\end{figure}

\begin{figure}
\includegraphics[angle=0.0,scale=.40]{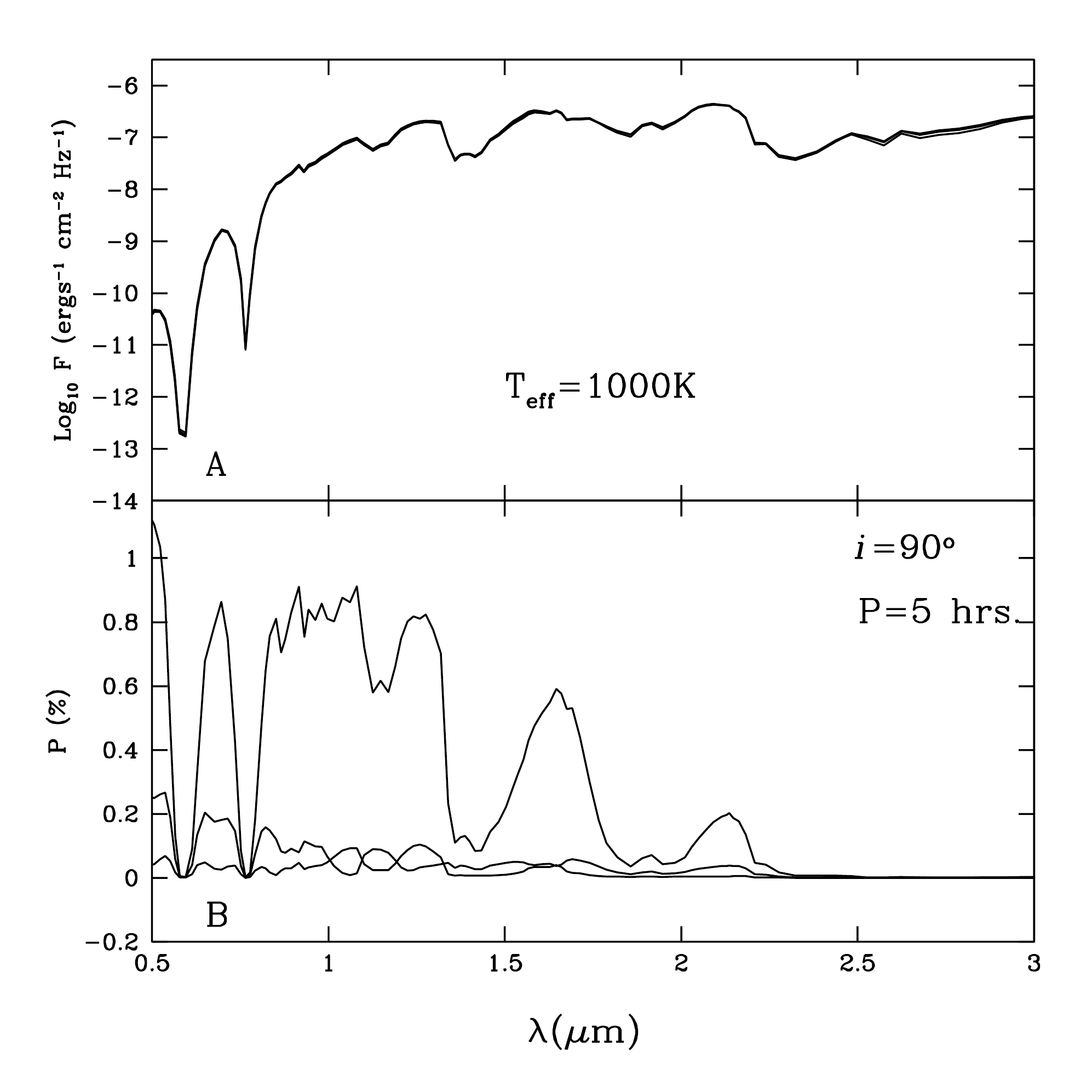}
\caption{The emergent flux (A) and the disk-integrated degree of linear
polarization (B) of non-irradiated exoplanets for a fixed
$T_{\rm eff}=1000$K but for different surface gravities. In (B), the solid
lines from top to bottom represent the polarization profiles for surface
gravity $g=30$, 56 and $100\,\rm m\,s^{-2}$ respectively. The difference in the
emergent flux (A) for a fixed effective temperature but surface gravity
varying over this range is not noticeable on this scale.  
\label{fig8}}
\end{figure}

\begin{figure}
\includegraphics[angle=0.0,scale=.40]{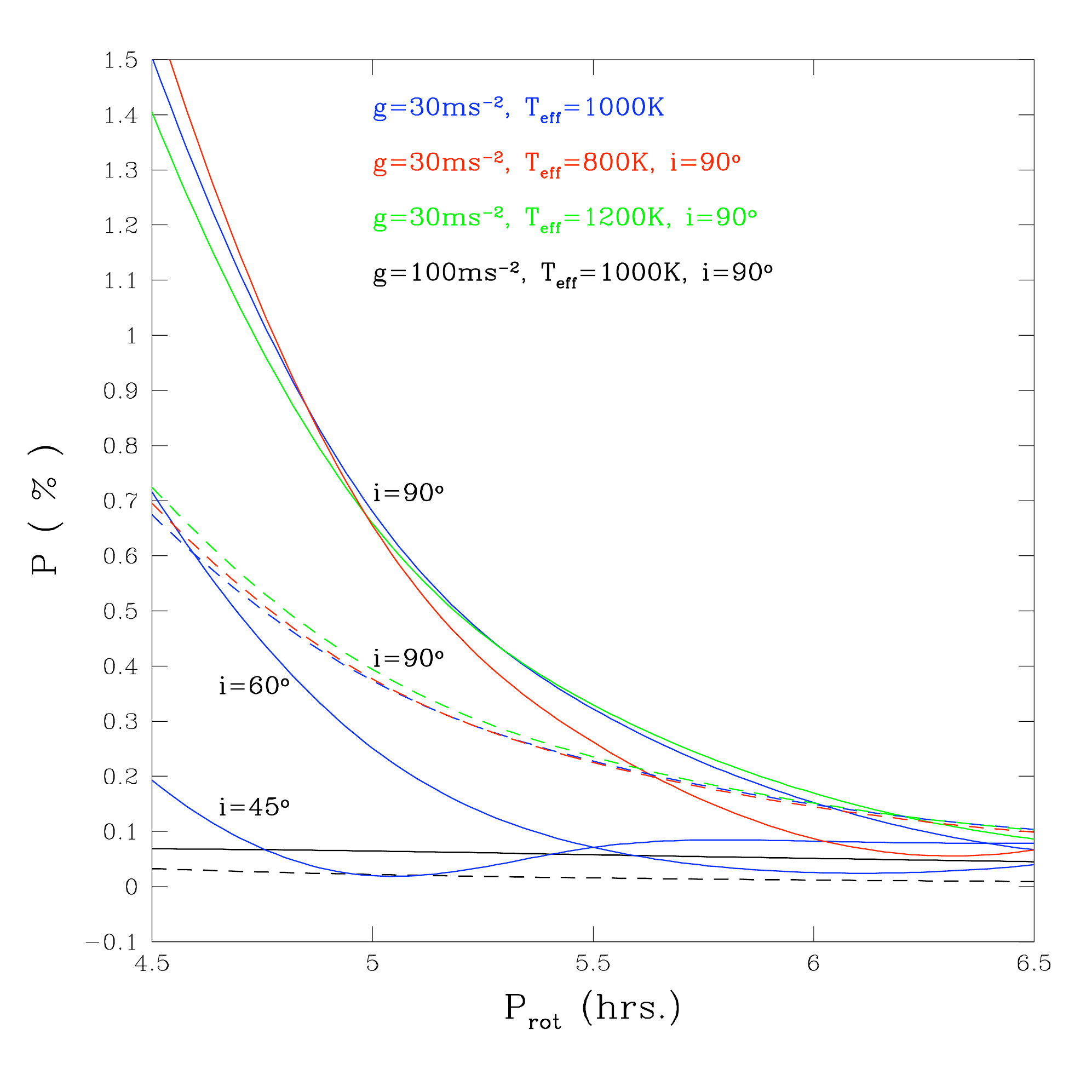}
\caption{
Scattering polarization profiles of non-irradiated exoplanets
with different rotational periods. The solid lines represent the  
percentage
degree of linear polarization in J-band while the broken lines represent
that in I-band. A variety of model cases are shown, all assume $f_{\rm sed}=2$.  Cases are shown for viewing angle $i=90^\circ$ (equator view) and $i=60$ and $45^\circ$.
\label{fig9}}
\end{figure}

\end{document}